\documentclass[aps,prx,twocolumn,groupedaddress,showpacs,floatfix]{revtex4}
\usepackage{mathptm}
\usepackage{dcolumn}
\usepackage{enumerate}
\usepackage{graphicx}

\begin{document}
\title{Localized versus itinerant states created by multiple oxygen vacancies in SrTiO$_3$}
\author{Harald O. Jeschke}
\affiliation{Institut f\"ur Theoretische Physik, Goethe-Universit\"at Frankfurt am Main, 60438 Frankfurt am Main, Germany}
\author{Juan Shen}
\affiliation{Institut f\"ur Theoretische Physik, Goethe-Universit\"at Frankfurt am Main, 60438 Frankfurt am Main, Germany}
\author{Roser Valent{\'\i}}
\affiliation{Institut f\"ur Theoretische Physik, Goethe-Universit\"at Frankfurt am Main, 60438 Frankfurt am Main, Germany}
\date{\today}

\begin{abstract}
  Oxygen vacancies in strontium titanate surfaces (SrTiO$_3$) have
  been linked to the presence of a two-dimensional electron gas with
  unique behavior.  We perform a detailed density functional theory
  study of the lattice and electronic structure of SrTiO$_3$ slabs
  with multiple oxygen vacancies, with a main focus on two vacancies
  near a titanium dioxide terminated SrTiO$_3$ surface. We conclude
  based on total energies that the two vacancies preferably inhabit
  the first two layers, {\it i.e.} they cluster vertically, while in
  the direction parallel to the surface, the vacancies show a weak
  tendency towards equal spacing. Analysis of the nonmagnetic
  electronic structure indicates that oxygen defects in the surface
  TiO$_2$ layer lead to population of Ti $t_{2g}$ states and thus
  itinerancy of the electrons donated by the oxygen vacancy. In
  contrast, electrons from subsurface oxygen vacancies populate Ti
  $e_g$ states and remain localized on the two Ti ions neighboring the
  vacancy. We find that both, the formation of a bound oxygen-vacancy
  state composed of hybridized Ti 3$e_g$ and 4$p$ states neighboring
  the oxygen vacancy as well as the elastic deformation after
  extracting oxygen contribute to the stabilization of the in-gap
  state.
\end{abstract}

\maketitle

{\it Introduction.}  The discovery of a two-dimensional electron gas
at the interface between SrTiO$_3$ (STO) and LaAlO$_3$ (LAO) in an
LAO/STO heterostructure by Ohtomo and Hwang~\cite{Ohtomo2004}
initiated intense research efforts~\cite{Mannhart2008,Huijben2009} on
these materials and unexpected phases at the interface like
superconductivity~\cite{Reyren2007} and
ferromagnetism~\cite{Brinkman2007} were reported.  However, there has
been some controversy on the mechanisms leading to the conducting
interface, with proposals ranging from electronic reconstruction as a
way to avoid a polar catastrophe~\cite{Okamoto2004} to various
mechanisms based on extrinsic defects like oxygen
vacancies~\cite{Siemons2007,Herranz2007} and site
disorder~\cite{Nakagawa2006,Li2014}. More recently, a metallic state
has also been discovered at the surfaces of freshly cleaved
SrTiO$_3$~\cite{Santander-Syro2011,Meevasana2011}
and KTaO$_3$~\cite{King2012}. In the case of pure SrTiO$_3$ surfaces,
the metallic state and the photoemission spectra can be well explained
with oxygen vacancies~\cite{Santander-Syro2011,Shen2012,Wang2014}.
Besides the spectral weight at the Fermi level, the presence of a peak
at about 1.3~eV below the Fermi level was also
reported~\cite{Santander-Syro2011}.  Aiura {\it et
  al.}~\cite{Aiura2002} observed in photoemission experiments for
lightly electron-doped SrTiO$_3$ under different oxygen pressure
conditions, that the peak at 1.3~eV appears to depend on the oxygen
defect density.  As pristine SrTiO$_3$ is a semiconductor with a large
band gap of $E_g=3.2$~eV. Therefore, creating a number of Ti $t_{2g}$
carriers and assuming a rigid band shift should lead to a
photoemission spectrum with a wide gap below the states near the Fermi
level. However, several
experiments~\cite{Courths1980,Aiura2002,Kim2009,Santander-Syro2011,Meevasana2011,Hatch2013}
show that the $E=-1.3$~eV feature is robust and reproducible but
sensitive to oxygen pressure. Understanding the nature and orbital
character of the $E=-1.3$~eV feature as well as the interplay between
localized and itinerant states created by the presence of oxygen
vacancies will be the main focus of our study.  In fact, the role of
oxygen vacancies is presently being intensively discussed in a wider
context of materials. For instance, oxygen vacancies have been
proposed to be responsible for the suppression of the metal-insulator
transition in VO$_2$~\cite{Jaewoo2013}, as well as for the electron
beam-induced growth of iron nanowires on TiO$_2$~\cite{Vollnhals2013},
to mention a few.  Therefore, getting a deeper microscopic
understanding of the role of oxygen vacancies in transition metal
oxides can further elucidate the mechanisms behind the above observed
phenomena.

There have been a number of previous theoretical efforts dealing with
oxygen vacancies in SrTiO$_3$. Hou and Terakura~\cite{Hou2010}
performed GGA+U calculations of single and double oxygen vacancies in
bulk SrTiO$_3$. Several calculations based on hybrid
functionals~\cite{Carrasco2006,Mitra2012} or LDA+U~\cite{Lin2012} find
an oxygen defect related in-gap state for SrTiO$_3$. Lin and
Demkov~\cite{Lin2013} use a three-orbital Hubbard orbital to study the
effect of electronic correlation on an oxygen vacancy in
SrTiO$_3$. Pavlenko {\it et al.}~\cite{Pavlenko2012} have analyzed the
orbital reconstruction at SrTiO$_3$/LaAlO$_3$ interfaces due to oxygen
vacancies within GGA+U. We will extend this existing work by (i)
focusing on SrTiO$_3$ surfaces and by (ii) using large supercells that
allow us to investigate two and three oxygen vacancies at realistic
defect densities.

\begin{figure}[ptb]
\includegraphics[width=0.47\textwidth]{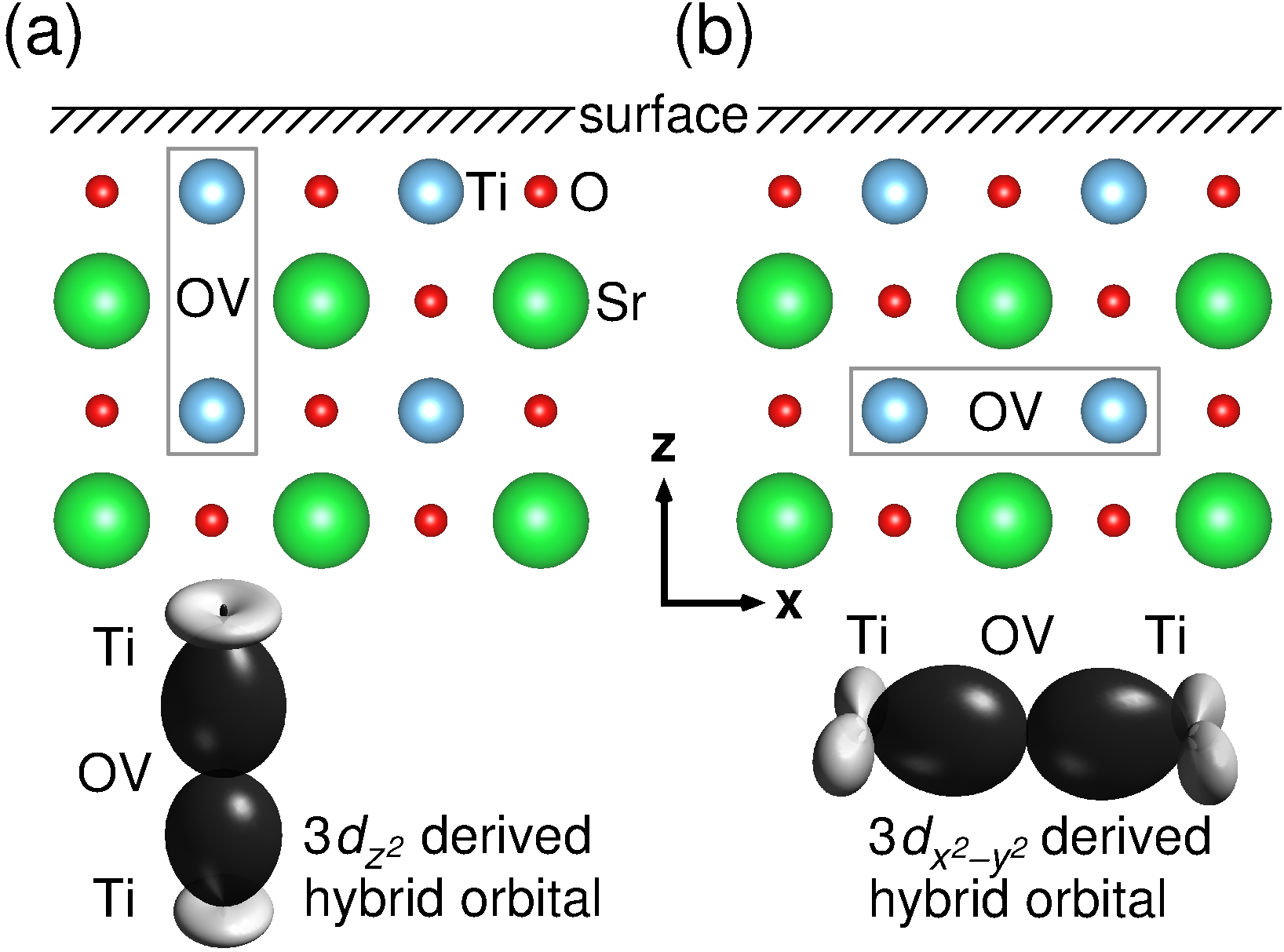}
\caption{(Color online) Schematic view of two inequivalent subsurface
  oxygen vacancy positions together with a sketch of the symmetry of
  the resulting hybrid orbitals on the Ti sites next to the oxygen
  vacancy.}
\label{fig:sto_scheme}
\end{figure}

In our study we show that (i) multiple subsurface oxygen defects are
energetically less favorable than configurations with at least one
defect in the TiO$_2$ surface layer. (ii) Vertically, defects show a
clear tendency to cluster; defect configurations with two oxygen
defects in the first two layers (surface TiO$_2$ and first subsurface
SrO layer) are clearly preferred over configurations with one or two
layers of vertical distance between the two vacancies. In contrast, in
the direction parallel to the surface, we find a tendency of vacancies
to distribute uniformly. (iii) Moreover, while the isolated surface
oxygen vacancy creates itinerant Ti $t_{2g}$ electrons, the subsurface
vacancy creates two localized states of Ti $e_g$ character in the two
adjacent Ti ions. The localized states have $3d_{z^2}$ character with
some $4p_z$ weight for oxygen vacancies in a subsurface SrO layer, and
$3d_{x^2-y^2}$ character with some $4p_x/p_y$ weight for vacancies in
a subsurface TiO$_2$ layer (see Fig.~\ref{fig:sto_scheme}).  Finally,
we also show that (iv) the precise condition for an in-gap state
produced from surface vacancies is the formation of a
TiO$_3$(vacancy)$_2$ cluster.

\begin{figure}[ptb]
\includegraphics[width=0.48\textwidth]{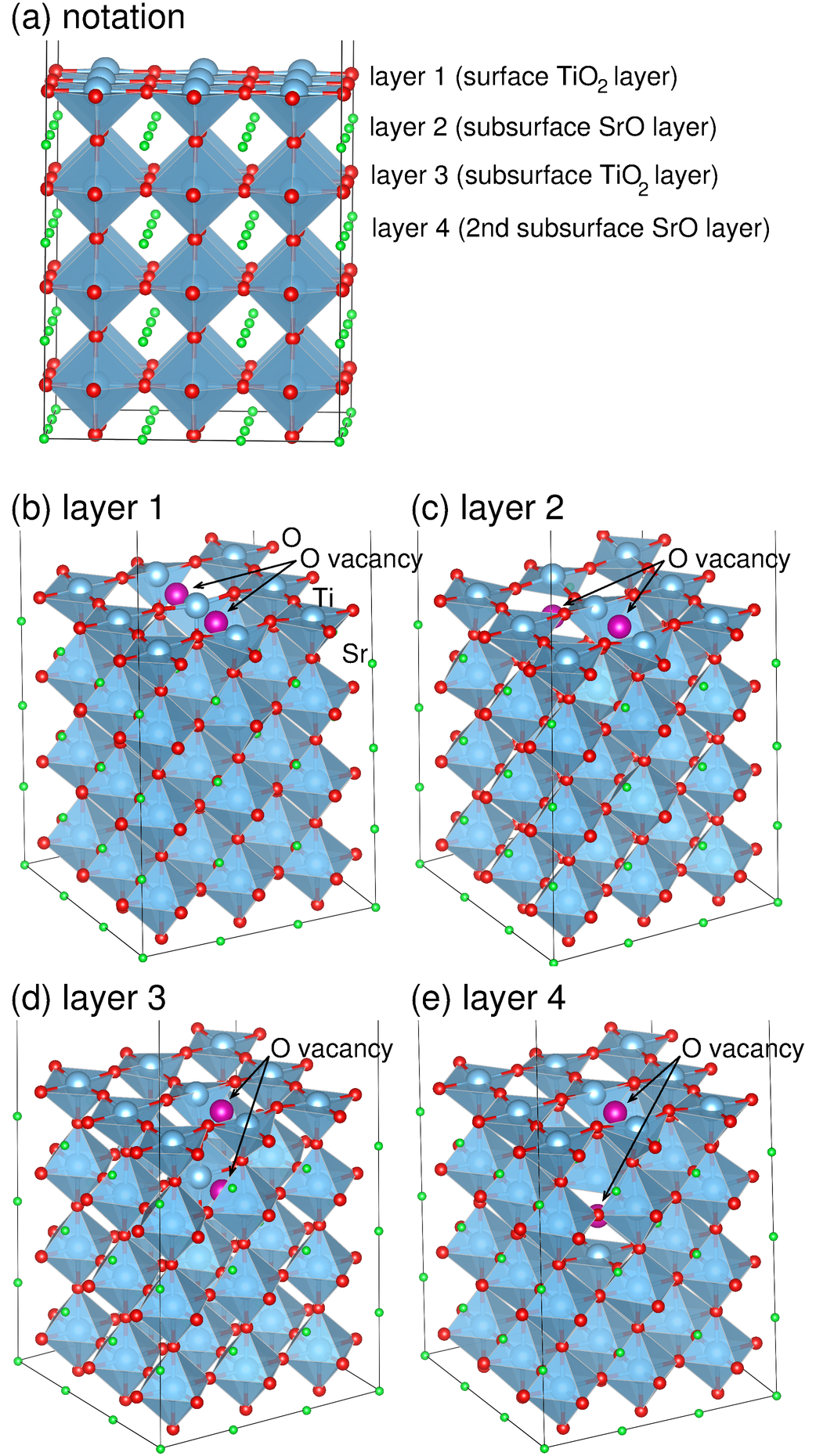}
\caption{(Color online) Examples of SrTiO$_3$ slab
  structures. $3\times 3\times 4$ perovskite units have been
  considered in the calculation and in (a) the notation used through
  the text is given. There are two oxygen vacancies: one is always in
  the surface TiO$_2$ layer (layer 1). Examples for the energetically
  most favorable positioning of the second oxygen vacancy in the
  surface layer (layer 1) is shown in (b), in the subsurface SrO layer
  (layer 2) in (c), in the first subsurface TiO$_2$ layer (layer 3) in
  (d) and in the second subsurface SrO layer (layer 4) in (e).}
\label{fig:structure}
\end{figure}

\begin{figure}[ptb]
\includegraphics[width=0.5\textwidth]{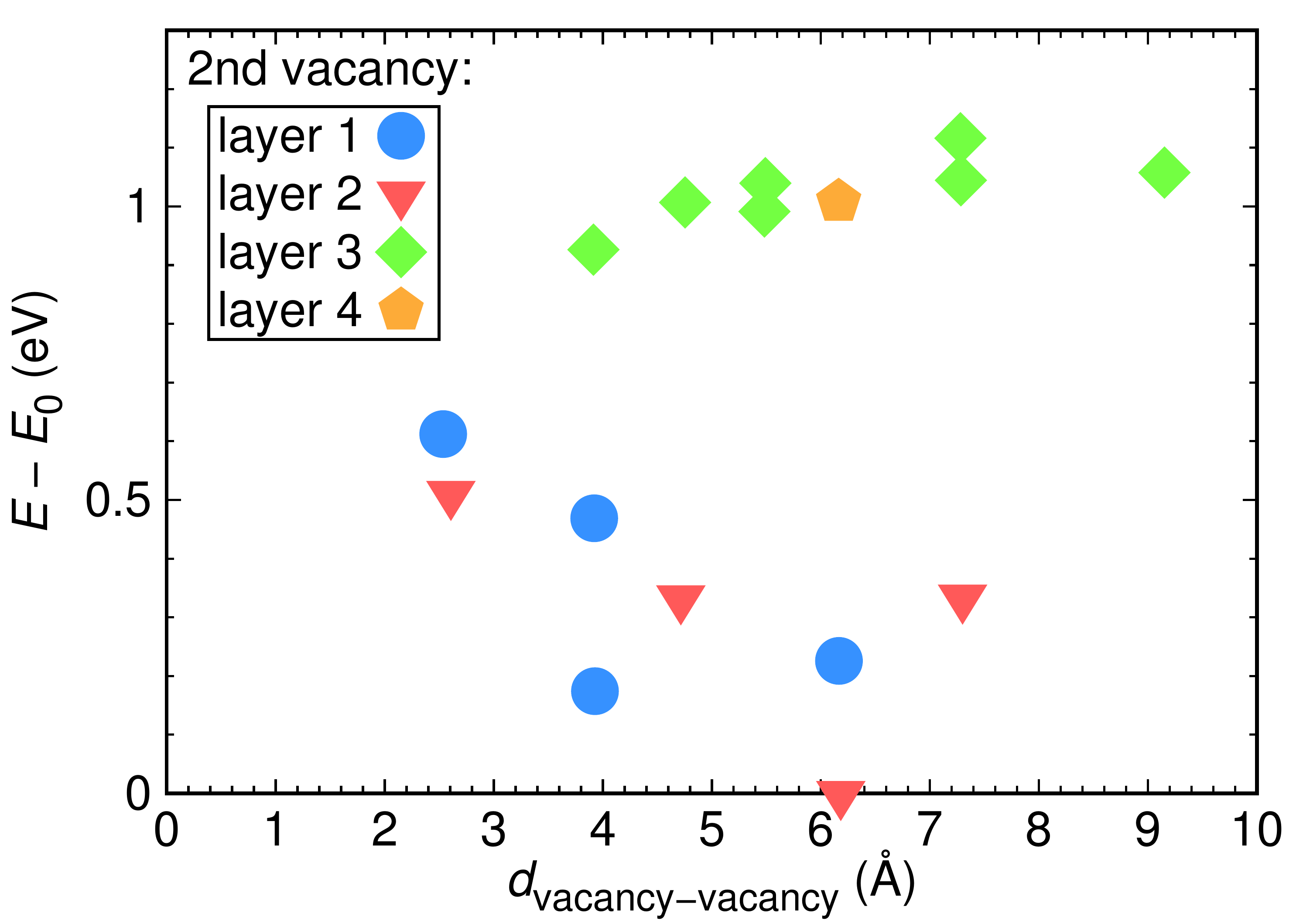}
\caption{(Color online) Total energies of SrTiO$_3$ slabs with two
  oxygen vacancies calculated within GGA+U. The first vacancy is
  always in the surface TiO$_2$ layer (layer 1). Energies are given as
  function of distance to the second vacancy which can be in the
  surface TiO$_2$ layer (layer 1) (circles), in the subsurface SrO
  layer (layer 2) (triangles), in the first subsurface TiO$_2$ layer
  (layer 3) (diamonds) or in the second subsurface SrO layer (layer 4)
  (pentagon).}
\label{fig:energies}
\end{figure}

{\it Method.}  In order to investigate the role of oxygen vacancies in
SrTiO$_3$, we performed density functional theory calculations for a
number of SrTiO$_3$ slabs with various configurations of oxygen
vacancies and analyzed the origin of the states appearing near the
Fermi level.  We have considered stoichiometric SrTiO$_3$ slabs with
(001) surfaces, as discussed in Ref.~\onlinecite{Shen2012}. Based on
our previous experience, we focus on $3 \times 3 \times 4$ supercells
with (a) TiO$_2$ and (b) SrO termination. We use the energetically
most favorable structures with a single vacancy in the TiO$_2$ or SrO
surface layer as starting point for structures with a second or even a
third oxygen defect.  We relax these structure candidates using the
Vienna \textit{ab initio} simulation package
(VASP)~\cite{Kresse1993,Kresse1996} with the projector augmented wave
(PAW) basis~\cite{Bloechl1994}. As it has been found that relaxations
with the generalized gradient approximation (GGA) tend to make the
octahedral environment of transition metal ions too
homogeneous~\cite{Foyevtsova2011} we use a GGA+U
functional~\cite{Liechtenstein1995} with literature values for
SrTiO$_3$~\cite{Okamoto2006} of $U=5$~eV and $J=0.64$~eV. We analyze
the electronic structure and total energy of the predicted slab
geometries using an all electron full potential local orbital
(FPLO)~\cite{Koepernik1999} basis.

\begin{figure}[ptb]
\includegraphics[width=0.5\textwidth]{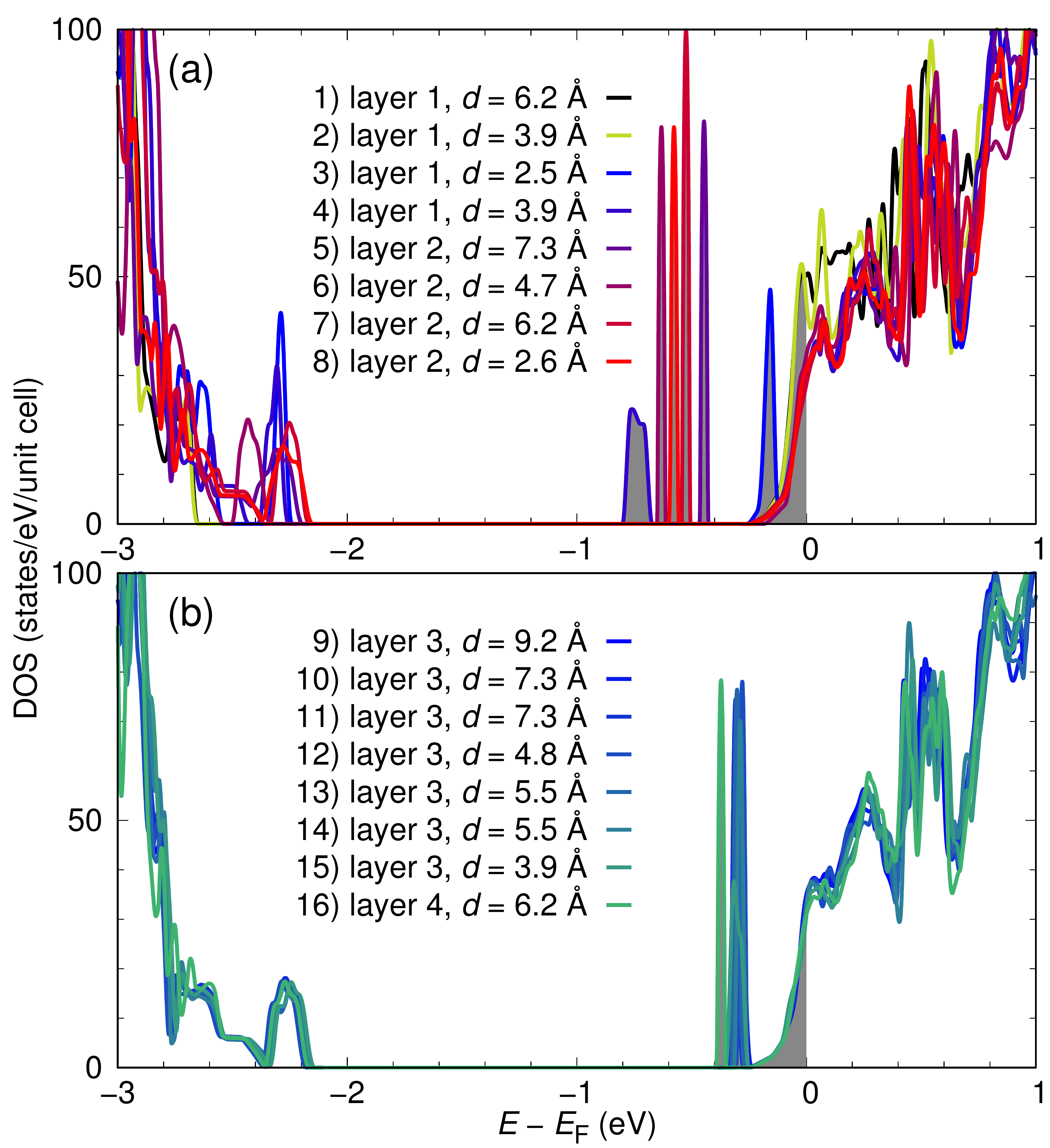}
\caption{(Color online) Densities of states for two oxygen vacancies
  in $3\times 3\times 4$ SrTiO$_3$ slabs calculated within GGA+U. The
  first vacancy is always in the surface TiO$_2$ layer. (a) Second
  vacancy in surface TiO$_2$ layer or in first subsurface SrO
  layer. All but the first two structures lead to an in-gap DOS
  peak. In the structure corresponding to DOS number 4, the two oxygen
  vacancies have a Ti in the middle, in contrast to DOS number 2.  (b)
  Second vacancy in first surface TiO$_2$ layer or in second
  subsurface SrO layer. Number 4 is the DOS for the structure in
  Fig.~\ref{fig:structure}~(a), number 8 corresponds to
  Fig.~\ref{fig:structure}~(b), number 15 to
  Fig.~\ref{fig:structure}~(c) and number 16 to
  Fig.~\ref{fig:structure}~(d).}
\label{fig:dostwo}
\end{figure}

\begin{figure}[ptb]
\includegraphics[width=0.5\textwidth]{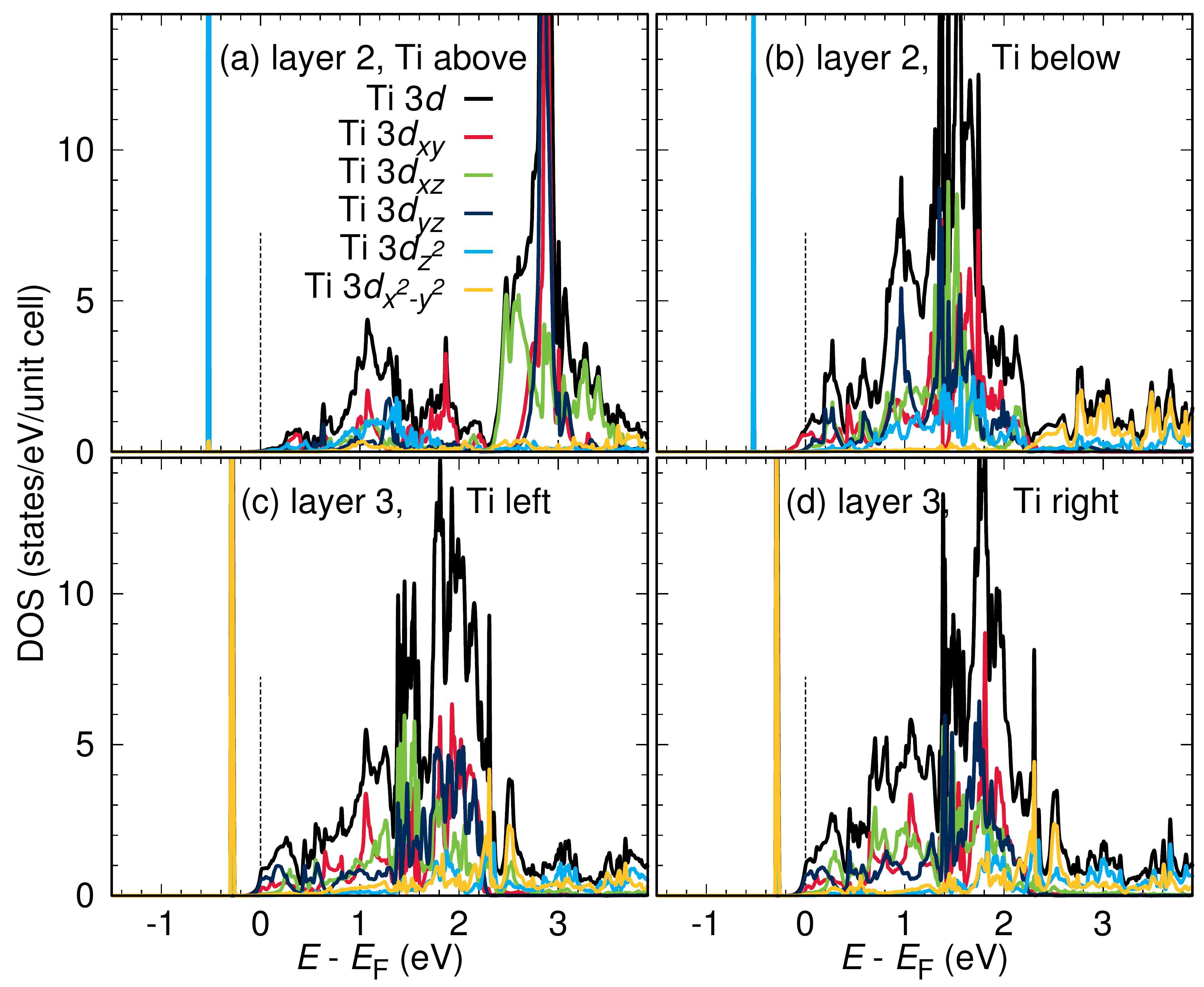}
\caption{(Color online) Partial densities of states for Ti ions
  neighboring oxygen vacancies. (a) and (b) Ti ions sitting above and
  below a vacancy in the first subsurface SrO layer. (c) and (d) Ti
  ions sitting right and left of a vacancy in the first subsurface
  TiO$_2$ layer. }
\label{fig:character}
\end{figure}

{\it Results.} In Figure~\ref{fig:structure}, we show examples of
SrTiO$_3$ supercells with two oxygen vacancies. They correspond to the
energetically most favorable configurations with the first vacancy in
the TiO$_2$ surface layer (layer 1) and the second vacancy in (a) the
surface layer (layer 1), (b) the first subsurface SrO layer (layer 2),
(c) the first subsurface TiO$_2$ layer (layer 3) or (d) the second
subsurface SrO layer (layer 4). An overview of the energetics is shown
in Figure~\ref{fig:energies}. Energies are given with respect to the
energy $E_0$ of the configuration drawn in
Figure~\ref{fig:structure}~(b) which turned out to be the optimum. We
find a clear trend: Defect configurations with one vacancy on the
surface (layer 1) and the second one in the first subsurface layer
(layer 2) are energetically more favorable than configurations where
the two oxygen vacancies are separated by one or two pristine
layers. This means that there is a clear tendency of oxygen vacancies
to cluster vertically near the surface of SrTiO$_3$. In the direction
parallel to the surface, however, the outcome of our simulations is
more complex. While configurations with both defects in the surface
TiO$_2$ layer (circles in Figure~\ref{fig:energies}) show a weak
tendency to cluster, the energetically most favorable corresponds to
distributing one defect in the first (TiO$_2$) and one defect in the
second (SrO) layer (triangle at $E=E_0$ in Figure~\ref{fig:energies})
with a maximal distance between the two vacancies within our
simulation cell. This result suggests a tendency to uniform
distribution of defects parallel to the surface. Turning to the
defects separated by a pristine SrO layer from the surface oxygen
defect (diamonds in Figure~\ref{fig:energies}), we observe a weak
preference of defects to lower vacancy-vacancy separation, {\it
  i.\,e.}  to cluster with the SrO spacer layer between the two
defects.

We now proceed with an analysis of the electronic structure of the two
oxygen vacancy configurations. Figure~\ref{fig:dostwo} shows the
density of states for all structures discussed before. Gray shading
indicates electronic states which are populated with electrons donated
by the oxygen vacancy. All investigated structures have Ti $t_{2g}$
weight near the Fermi level. A detailed analysis of this weight shows
that a large amount of Ti ions in the supercell contribute to it and
the corresponding bands are dispersive, indicating that these
electrons are itinerant. We have shown in Ref.~\onlinecite{Shen2012}
that this is a result of structural relaxation; if all ions are kept
in the ideal perovskite position upon creation of an oxygen vacancy,
only Ti $t_{2g}$ orbitals next to the vacancy are occupied, and an
unphysical localized $t_{2g}$ electron density is created.  As a
second important feature, all structures in
Figure~\ref{fig:dostwo}~(b) and all except the first one in
Figure~\ref{fig:dostwo}~(a) also show sharp in-gap states; these are
states typically created by subsurface oxygen defects and localized on
the two Ti ions adjacent to the defect. These states have Ti $e_g$
character with small admixture of $4p$ states, and they clearly fall
into two categories: Ti $d_{z^2}$ states created by vacancies in a SrO
layer, and Ti $d_{x^2-y^2}$ states produced by vacancies in subsurface
TiO$_2$ layers. This orbital occupancy is due to the fact that in the
case of an oxygen defect in a SrO layer, the Ti ions neighboring the
defect are above and below the defect where the vertical axis
corresponds to the $z$ axis in the orbital projection (see
Fig.~\ref{fig:sto_scheme} (a)).  In the case of a subsurface defect in
a TiO$_2$ layer, the two neighboring Ti ions sit at half a lattice
spacing either along $x$ or along $y$ direction with respect to the
vacancy (see Fig.~\ref{fig:sto_scheme} (b)).
Figure~\ref{fig:character} (a) illustrates the orbital distribution of
the defect configuration shown in Figure~\ref{fig:structure}~(b) where
one vacancy is on the TiO$_2$ surface (layer 1) and the second vacancy
is on the subsurface SrO layer (layer 2).  Figure~\ref{fig:character}
(b) displays the orbital distribution for a representative example of
one vacancy on the TiO$_2$ surface (layer 1) and the second one on the
TiO$_2$ subsurface (layer 3), as in Figure~\ref{fig:structure}~(c).

In-gap states are also present when oxygen vacancies cluster at the
surface TiO$_2$ layer (cases 3 and 4 in Fig.~\ref{fig:dostwo}). Our
calculations show that the precise condition for such in-gap states
produced from surface vacancies is the formation of a
TiO$_3$(vacancy)$_2$ cluster. In fact, the energetically most
favorable configuration with two vacancies in the surface TiO$_2$
layer (see Figure~\ref{fig:structure}~(a)) is of this type. On the
other hand, well separated oxygen vacancies in the surface TiO$_2$
layer which form TiO$_4$(vacancy) clusters only (cases 1 and 2 in
Fig.~\ref{fig:dostwo}) lead to an itinerant 2D electron gas of Ti
$t_{2g}$ electrons but no in-gap states.

\begin{figure}[ptb] 
\includegraphics[width=0.5\textwidth]{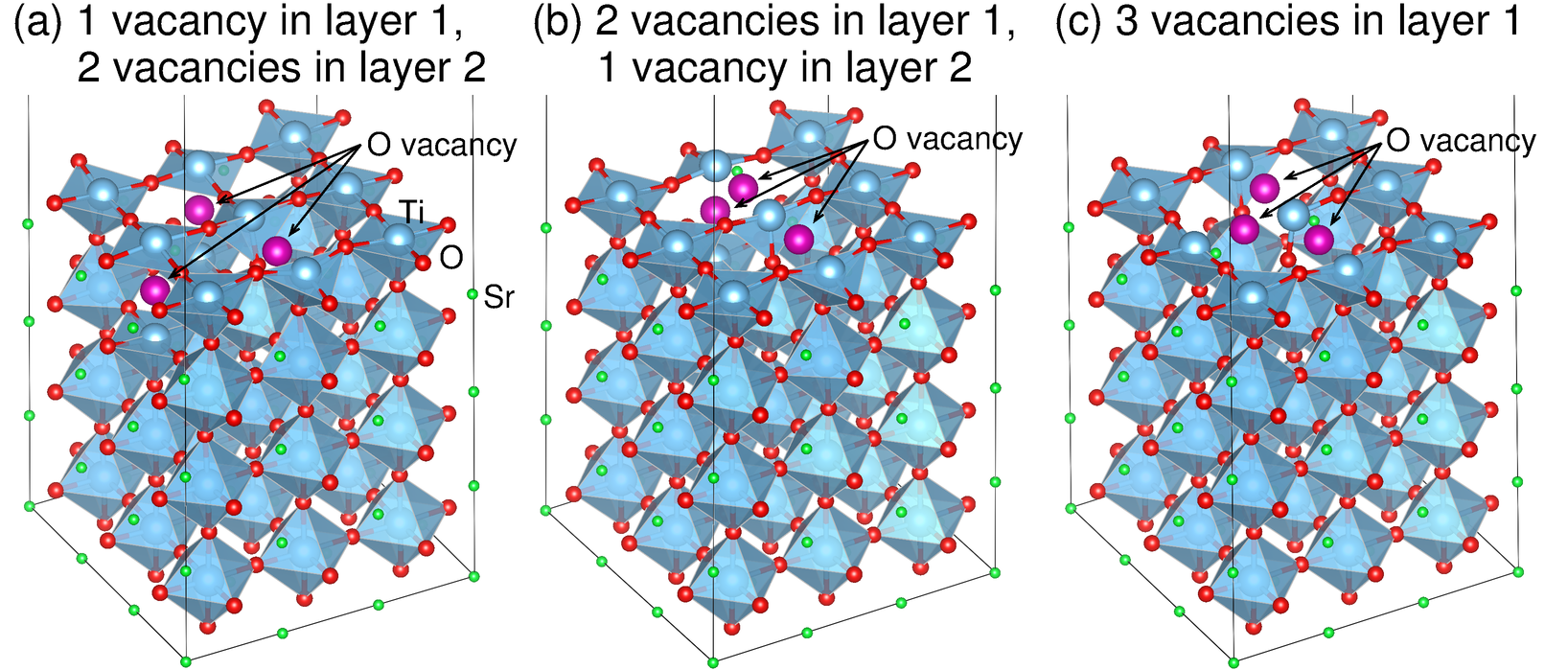}
\caption{(Color online) Three examples of three oxygen vacancy
  configurations in $3\times 3\times 4$ SrTiO$_3$ slabs. }
\label{fig:structthree}
\end{figure}

\begin{figure}[ptb] 
\includegraphics[width=0.5\textwidth]{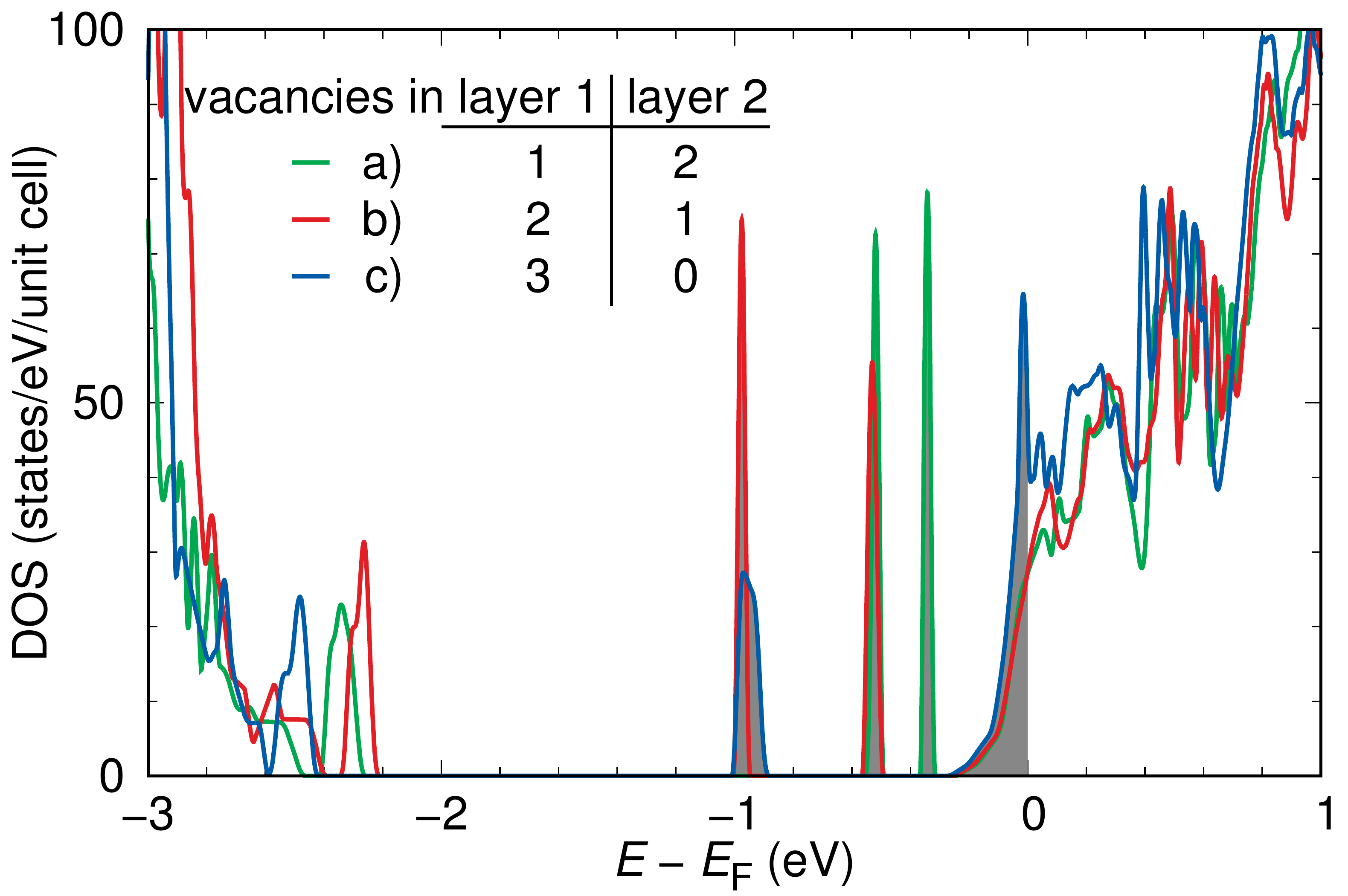}
\caption{(Color online) Densities of states for three examples of
  three oxygen vacancy configurations in $3\times 3\times 4$ SrTiO$_3$
  slabs calculated within GGA+U. }
\label{fig:dosthree}
\end{figure}

In order to further test the distribution of oxygen-vacancy-induced
extra charge, we also calculated the electronic properties of $3\times
3\times 4$ SrTiO$_3$ slabs with three oxygen vacancy configurations as
shown in Fig.~\ref{fig:structthree}. In all three cases in-gap states
appear below the Fermi level since some oxygen vacancies are either
below the TiO$_2$ surface layer or they are clustered around a Ti on
the surface.  Comparison of the two-vacancy with the three-vacancy
cases shows that the in-gap weight is proportional to the oxygen
vacancy density.  This observation is in qualitative agreement with
experiment~\cite{Aiura2002} but it should be investigated further.

{\it Discussion.}  Analysis of the previous LDA+U results allows us to
draw some important conclusions regarding the role of oxygen vacancies
in SrTiO$_3$.  (1) If oxygen vacancies are only on the surface and
well separated from each other, the two electrons per vacancy
contribute {\it only} to the conduction band and no localized in-gap
states are formed independently of the $U$ value considered in the
LDA+U calculations. Only when oxygen vacancies cluster on the surface,
assuming TiO$_3$(vacancy)$_2$ configurations, or are positioned in
subsurface layers do we observe the formation of in-gap states coming
from the hybridized $3e_g$ with $4p$ states from the Ti neighboring
the vacancy. This is in contrast to a recent study by Lin {\it et
  al.}~\cite{Lin2013} where it was suggested that the oxygen-vacancy
induced in-gap state traps at most one electron from the oxygen
vacancy while the second electron contributes to the conduction.  (2)
The energy ordering of the different vacancies configurations with
presence of in-gap states can be attributed to two effects: (i) the
gain in energy due to the formation of a bound oxygen-vacancy state
(in-gap state) composed of the hybridized Ti $e_g$ and $4p$ states
neighboring the vacancy as well as (ii) the gain in elastic energy due
to the lattice deformation after extracting oxygen. In fact,
calculations of total energies of relaxed versus unrelaxed slab
structures point to a significant contribution of the second effect
that should be considered together with the formation of the bound
state. Moreover, the lattice deformation is important in the formation
of an itinerant 2D electron gas due to surface oxygen vacancies. (3)
The weight of the in-gap state scales with the oxygen vacancy
concentration in agreement with photoemission experiments. (4) The
formation energy of an oxygen vacancy in SrTiO$_3$ is about 7.7~eV for
a single vacancy and 4.8~eV per vacancy for two and three
vacancies. From our present calculations we can only speculate about
possible formation mechanisms of such vacancies.  Certainly the
exposure to energetic photons in photoemission experiments is a
possible cause.

In summary, by considering different configurations of oxygen
vacancies in SrTiO$_3$ and subsequent analysis of the energetics and
electronic properties via extensive DFT calculations we can explain
the origin of observed in-gap states as well as conduction electrons
in photoemission experiments on SrTiO$_3$ surfaces and provide
predictions for the behavior of a finite concentration of oxygen
vacancies in SrTiO$_3$.

\acknowledgments

We thank Ralph Claessen, Michael Sing, Andres Santander-Syro, Marc
Gabay and Marcelo Rozenberg for useful discussions and gratefully
acknowledge financial support by the Deutsche Forschungsgemeinschaft
(DFG) through grant FOR 1346. The generous allotment of computer time
by CSC-Frankfurt and LOEWE-CSC is also gratefully acknowledged.

\end{document}